\documentstyle[aaspp4,12pt,psfig]{article}
\lefthead{Bower et al.}
\righthead{Circular Polarization of Sgr~A*}
\begin{document}

\newcommand\degd{\ifmmode^{\circ}\!\!\!.\,\else$^{\circ}\!\!\!.\,$\fi}
\newcommand{\etal}{{\it et al.\ }}
\newcommand{\uv}{(u,v)}
\newcommand{\rdm}{{\rm\ rad\ m^{-2}}}

\title{Detection of Circular Polarization in the Galactic
Center Black Hole Candidate Sagittarius A* }

\author{Geoffrey C. Bower\altaffilmark{1}, 
Heino Falcke\altaffilmark{2,3} \&
Donald C. Backer\altaffilmark{4} }

\altaffiltext{1}{National Radio Astronomy Observatory, P.O. Box O, 1003 
Lopezville, Socorro, NM 87801; gbower@nrao.edu} 
\altaffiltext{2}{Max Planck Institut f\"{u}r Radioastronomie, Auf dem 
H\"{u}gel 69, D 53121 Bonn Germany; hfalcke@mpifr-bonn.mpg.de} 
\altaffiltext{3}{Steward Observatory, The University of Arizona,
Tucson, AZ 85721} 
\altaffiltext{4}{Astronomy Department \& Radio Astronomy Laboratory, 
University of California, Berkeley, CA 94720; dbacker@astro.berkeley.edu}

\begin{abstract}

We report here the detection of circular polarization in the 
Galactic Center black hole candidate, Sagittarius A*.  
The detection was made at 4.8 GHz and 8.4 GHz with the 
Very Large Array.  We find that the fractional circular polarization 
at 4.8 GHz is $m_c=-0.36 \pm 0.05\%$ and that the spectral index 
of the circular polarization is $\alpha=-0.6 \pm 0.3$ ($m_c \propto
\nu^{\alpha}$).  The systematic error in $m_c$ is less than 
0.04\% at both frequencies.  In light of our recent lower limits on the
linear polarization in Sgr A*, this detection is difficult to
interpret with standard models.  
We consider briefly whether scattering mechanisms could
produce the observed polarization.  Detailed modeling of the
source and the scattering medium is necessary.
We propose a simple model in which low energy
electrons reduce  linear polarization through Faraday
depolarization and convert linear polarization into circular polarization.
Circular polarization may represent a significant new parameter
for studying the obscured centimeter wavelength radio source in
Sgr A*.
\end{abstract}


\section{Introduction}

The compact radio source in the Galactic Center, Sagittarius~A*,
is the best and closest candidate for a supermassive black hole in the
center of a galaxy (Maoz \markcite{maoz98} 1998).
The source Sgr A* is
positionally coincident with a $\sim 2.6 \times 10^6 M_{\sun}$ dark
mass (Genzel \etal \markcite{genze97} 1997, Ghez \etal \markcite{ghez98} 1998). 
Very long baseline interferometry (VLBI) has shown that
this source has a scale less than 1 AU and a brightness temperature in
excess of $10^9 {\rm\ K}$ (Rogers \etal \markcite{roger94} 1994, 
Bower \& Backer \markcite{bower98} 1998, 
Lo \etal \markcite{lo98} 1998, Krichbaum \etal 1998).  
Long-term studies of Sgr~A*
indicate that the source shows no motion with respect to the center of
the Galaxy (Backer \& Sramek \markcite{backe99} 1999,  
Reid, Readhead, Vermeulen, \& Treuhaft \markcite{reid99} 1999). 
For these reasons, it is inferred that Sgr~A* is a
supermassive black hole with a synchrotron emission region fed
through accretion (Melia \markcite{melia94} 1994, 
Narayan \etal \markcite{naray98} 1998, Falcke, Mannheim \& 
Biermann \markcite{falck93} 1993, Mahadevan \markcite{mahad98} 1998).  
In this view, Sgr A* is a weak active galactic nucleus (AGN).
However, strong interstellar scattering of the
radiation along the line of sight has been shown to broaden the image
of Sgr~A* at radio through millimeter wavelengths (e.g., 
Lo \etal \markcite{lo98} 1998, Frail \etal \markcite{frail94} 1994).
As a consequence, VLBI observations have not convincingly demonstrated
the existence of source structure that would be an important
diagnostic of physical processes.

Polarization has proved to be an important tool in the study of AGN.
Studies of linear polarization, which is typically on the order of a
few percent or less of the total intensity, have confirmed that the
emission process is synchrotron radiation and demonstrated that shocks
align magnetic fields in a collimated jet, leading to correlated 
variability in the total and polarized intensity (Hughes, Aller \& Aller
\markcite{hughe85} 1985, Marscher \& Gear \markcite{marsc85} 1985).
Circular polarization, on the other hand, is
less well understood in AGN.  Typically, the degree of circular
polarization is $m_c < 0.1\%$ with only a few cases where $m_c$
approaches $0.5\%$ (Weiler \& de Pater \markcite{weile83} 1983).
The degree of circular polarization usually
peaks near 1.4 GHz  and decreases strongly with increasing frequency.

Recently, VLBI imaging of 3C 279 has found 
$m_{\rm c}\simeq1\%$ 
in an individual radio component with a fractional linear polarization
of 10\% (Wardle \etal \markcite{wardl98} 1998).
The integrated circular
polarization, however, is less than 0.5\%. 
The circular polarization is probably produced through the conversion
of linear to circular polarization by low-energy electrons in the 
synchrotron source.  This process is also known as repolarization 
(Pacholczyk \markcite{pacho77} 1977).

In recent work we have
shown that the linear polarization of Sgr~A* from centimeter to
millimeter wavelengths is extremely low. 
Linear polarization was not detected in a spectro-polarimetric
experiment with an upper limit of 0.2\% for
rotation measures as large as $10^7 \rdm$ at 8.4 GHz 
(Bower \etal \markcite{bower99a} 1999a).
More recently, we have found that linear polarization is less than 0.2\%
at 22 GHz and less than $\sim1\%$ at 86 GHz 
(Bower \etal \markcite{bower99b} 1999b).
Interstellar depolarization is very unlikely within the 
parameter space covered by these observations.  
Given these stringent limits on linear polarization, the
presence of circular polarization is not expected.  Nevertheless, we
have detected circular polarization at a surprisingly
high level.

\section{Observations and Results}

We observed Sgr~A* with the Very Large Array (VLA) of the
National Radio Astronomy Observatory\footnote{The
National Radio Astronomy Observatory is a facility
of the National Science Foundation operated under cooperative agreement
by Associated Universities, Inc.} 
in its A
configuration on 10, 18 and 24 April 1998 at 4.8 GHz and 8.4 GHz with
a bandwidth of 50 MHz in two intermediate frequency (IF) bands
in both right circular polarization (RCP) and
left circular polarization (LCP).  
At 4.8 GHz, we cycled rapidly
between 2.5 minute scans on Sgr~A* and the nearby calibrators
B1737-294, B1742-283 and B1745-291.  Every hour the calibrators
B1741-038 and B1748-253 were observed.  These observations covered a
range of four hours.  A similar approach was used at 8.4 GHz over only
a single hour.  Absolute fluxes were calibrated with the source
3C~286.  Time-dependent amplitude calibration of the array was
performed through self-calibration of the compact and bright quasar
B1741-038.  Each source was phase self-calibrated.  All presented results
were done using baselines greater than 100$k\lambda$ in order to resolve out
large scale structure in the Galactic Center (e.g., Yusef-Zadeh, Roberts \&
Biretta \markcite{yusef98} 1998).

Detection of circular polarization with an
array with circularly-polarized feeds requires careful calibration
and is subject to a variety of errors.  The requirements are more complex
for Sgr A*, which is located in a significantly confused region.
Below, we summarize the errors that arise for standard circular polarization
measurements and show that they agree with the measured values for
the calibrators.  Following that, we demonstrate that the background radiation
for Sgr A* is not responsible for the measured signal.

The Stokes parameter $V$ is formed from the 
difference of the left-
and right-handed parallel polarization correlated visibilities, $LL$
and $RR$.  This difference is sensitive to amplitude calibration
errors.  Gain variations with time
were measured to be less than $\sim 0.3\%$ for all antennas.  
Averaged over independent gain measurements at 
all antennas this implies a calibration error of $\sim 0.02\%$
and $\sim 0.06\%$ at 4.8 and 8.4 GHz, respectively.

Beam squint may also introduce false circular polarization for objects
off-axis (Chu \& Turrin \markcite{chu73} 1973).  
Beam squint is due to the slightly displaced RCP and LCP beams
in an offset reflector antenna.  In the case of the VLA antenna geometry, 
the offset is 4.6\% of 
the primary beam FWHM (Condon \etal \markcite{condo98} 1998).  
At 4.8 GHz, this corresponds to $0.50^{\prime}$
within a primary beam of $10.82^{\prime}$.  All of our sources were observed
on the primary axis with a precision of less than $1^{\prime\prime}$.
However, pointing errors for 
individual antennas can be as large as $10^{\prime\prime}$.  This will lead
to a false circular polarization of $\sim 0.5\%$ in a single observation
on a single antenna.  Averaging over multiple observations with
the entire array produces a false circular polarization $\sim 0.03\%$.
At 8.4 GHz, the expected false circular polarization is $\sim 0.10\%$.

Second-order polarization leakage effects may produce false circular
polarization, as well.  For an array such as the VLA with polarization
leakage terms on the order of 1\%, weakly linearly polarized sources 
will produce false circular polarization on the order of a few times
0.01\%.  We performed our analysis with and without polarization
leakage correction and found no difference in the results.

Finally, false circular polarization may appear through interference.
We analyzed each of the 12 2.5-minute scans for Sgr~A* at 4.8 GHz
in the 10 April 1998 data independently 
and found no dependence on time in Stokes $V$.  The typical rms image noise in
each scan was 375 $\mu{\rm Jy}\ {\rm beam^{-1}}$, or 0.07\%.  Assuming that the
circular polarization did not change, we find $\chi^2=13.0$
for 11 degrees of freedom.  Interference may also have a frequency
dependence.  We measured Stokes $V$ for the separate IF bands for Sgr A*
at 4.8 GHz in the 10 April data to be -0.43\% and -0.29\%, which
is consistent with no frequency dependence for the error
determined below.

Adding in quadrature errors from amplitude calibration, beam squint
and polarization leakage gives total errors for a measurement
corresponding to a single day 
of $\sim 0.05\%$ and
$\sim 0.12\%$ at 4.8 and 8.4 GHz, respectively.  We estimate the
error in the average measurements to be 0.03\% and 0.07\%.
These values
are very similar to the values observed for the calibrator sources.  
We estimate the error in the mean for Sgr A* from the variance of the three
measurements to be 0.05\% and 0.06\% at the two
frequencies, respectively.  These estimates are very close to the
error determined above, which suggests that we are accounting for
most sources of error.  However, we now detail additional sources
of error and how we eliminate them independently.

Sgr A* may 
have additional errors because of the presence of significant
extended and compact structure in the Galactic Center region.  
We eliminate the effect of this structure on the
correlated visibilities by comparing
the given results with those obtained using all baselines,
only baselines 
greater than $300 k\lambda$, and only baselines less than $300 k\lambda$.  
We find at 4.8 GHz for 10 April 1998
$m_c=-0.36\%$, -0.34\%, and -0.37\%, respectively.

A source in the beam sidelobes could also introduce a false signal 
if it is circularly-polarized or if the sidelobe response is 
circularly-polarized.  The signature of this effect would include
time-dependence and possibly frequency-dependence as the sidelobe
swept over the source.  But we have demonstrated that these effects
do not exist at our sensitivity levels.

Finally, non-linear response of the detectors to the 
extended structure will introduce a systematic offset in circular
polarization between the amplitude calibrator and Sgr A*.
The detector response is linear to better than $1\%$.
The shift in system temperature between the calibrator and Sgr A*
is from 25 to 35 K at 4.8 GHz, implying a systematic error of $\la 0.3\%$
per antenna per polarization.  Averaging over both polarizations and
all antennas, one expects a systematic offset of unknown sign
 in Stokes $V$ due to the background radiation of $\la 0.04\%$.  
The magnitude of this effect at 8.4 GHz is also $\la 0.04\%$.

We show in Figure~1 the Stokes $V$ image of Sgr A* 
at 4.8 GHz from
10 April 1998.  The peak at -1.8 mJy is more than 20 times the
noise level of 68 $\mu {\rm Jy}\ {\rm beam^{-1}}$.  
For the same epoch, the calibrator B1748-253 has a peak
flux of 262 $\mu {\rm Jy}$ in a map with a noise level of 
58 $\mu {\rm Jy}\ {\rm beam^{-1}}$.
The total intensities of Sgr~A* and B1748-253 are 0.525 and 0.488 Jy
at the time of detection, respectively.

We summarize in Table~1 the total and circularly polarized intensities
for all sources at 4.8 and 8.4 GHz in our experiment.  At 4.8 GHz, the
calibrators all have mean circular polarizations less than 0.02\% with
the exception of B1737-294 which is dominated by the thermal noise of
$\sim 60\ \mu {\rm Jy}$.  The detection of circular polarization in
Sgr~A* is very certain at 4.8 GHz.  At 8.4 GHz, the greater thermal
noise and the less accurate calibration make the detection of circular
polarization for
Sgr~A* in each observation less certain.  However, the average result
firmly demonstrates detection.  The average circular polarization
 flux is 16 times that of B1748-253.

We find $m_c=-0.36 \pm 0.05\%$ and $m_c=-0.26 \pm 0.06\%$ at 4.8 and
8.4 GHz, respectively.  The errors are estimated from the variance 
of the three separate measurements for each frequency.
These errors set an upper limit to the variability, as well.
The average spectral index of the fractional circular polarization
is $\alpha=-0.6 \pm 0.3$ for $m_c \propto \nu^\alpha$.  
The error in $\alpha$ is less than that expected from the errors in
$m_c$.  This is due to the fact that variations in $m_c$ between
epochs appear to be due to systematic errors that are common to both
frequencies.

\section{Mechanisms for the Production of Circular Polarization}

While the detection of circular polarization for Sgr A* is in itself
an unexpected result, two additional properties set this source apart
from other radio cores and make the result difficult to understand:
the fact that circular polarization exceeds linear by more than a factor
of two and the flatness of the circularly polarized spectrum.

We have established previously that the interstellar scattering does
not depolarize any intrinsic linearly polarized emission from 
Sgr A* (Bower \etal 1999a, 1999b).
However, the sub-parsec accretion region of Sgr A* may have
very large rotation measures (Bower \etal 1999a).  This region may
depolarize an intrinsic linearly polarized signal without interfering
with the circularly polarized radiation.  Detailed modeling of the
accretion region may be able to address these issues (e.g., Melia \&
Coker \markcite{melia99} 1999).

We consider now whether the circular polarization is 
produced not in the source, but in the intervening scattering screen.  A
birefringent scattering medium may produce scintillating circular
polarization from an unpolarized background source.
This effect has been studied in detail 
(Macquart \& Melrose \markcite{macqu99} 1999).
This requires a scattering region with a fluctuating
rotation measure gradient.  Such a mechanism is
appealing due to the strong scattering medium and the strong observed
gradients in RM in the GC region (Yusef-Zadeh, Wardle \& Parastaran 
\markcite{yusef97} 1997). 
The fact that
the scattered image of Sgr A* itself is anisotropic might also
indicate an anisotropic scattering medium.  However, the diffractive effect
has $\alpha=-4$ or steeper, which is not consistent with the
measured spectral index. Further calculations of this relatively
unexplored issue should show whether such a scattering model could
nevertheless be made consistent with the observations.

Alternatively, we can ask whether the conversion mechanism
or intrinsic synchrotron circular polarization could be at work in Sgr A* 
(Pacholczyk \markcite{pacho77} 1977, Jones \& O'Dell \markcite{jones77} 1977). 
Here, the main problem is the low level of linear
polarization. Magnetic field reversals would reduce linear
polarization, but most likely would affect circular polarization in
the same way (Wilson \& Weiler \markcite{wilso97} 1997).
However, one important factor in the relative level of linear to
circular polarization is the electron energy distribution, since
low-energy electrons (with Lorentz factors less than 100)
can lead to Faraday-depolarization of linear
and/or conversion of linear to circular polarization.

As model calculations show (Jones \& O`Dell, Fig.1) circular
polarization of a radio component peaks near the self-absorption
frequency $\nu_{\rm ssa}$, where linear polarization drops to a
minimum. In typical radio core components the strongest contribution
to linear polarization therefore comes from the optically-thin
power law part of its spectrum. However, in Sgr A* such a power law is
most likely absent or already ends at a frequency $\nu_{\rm
max}\sim\nu_{\rm ssa}$, as indicated by the steep high-frequency
cut-off in its spectrum towards the infrared 
(Serabyn \etal \markcite{serab97} 1997,
Falcke \etal \markcite{falck98} 1998).

Such a situation has not yet been considered in synchrotron
propagation calculations involving conversion. However, it
may result in a high $m_{\rm c}$-to-$m_{\rm l}$ 
ratio if $\nu_{\rm max}\sim\nu_{\rm ssa}$ is true for the
electrons that produce the low frequency spectrum.
For example, Jones \& O'Dell show
that for a power law of electron energies with characteristic frequencies
$\nu_{\rm min}$ extending at least a factor thirty below $\nu_{\rm
ssa}$, circular polarization around $\nu_{\rm ssa}$ can 
exceed $m_{\rm l}$. With the absence of any higher frequency
emission, linear polarization could be quenched.
On the other hand, a narrow distribution
with $\nu_{\rm min}\sim\nu_{\rm ssa}\sim\nu_{\rm max}$ would again
lead to significant linear polarization even at the self-absorption
frequency (Jones \& Hardee \markcite{jones79} 1979).

We consider here a simple synchrotron model of Sgr A*,
where the flux density at 5 GHz is produced in a spherical component 
by a flat electron distribution with a power law index of $p=1$
ranging over electron energies that correspond to the characteristic
frequencies
$\nu_{\rm max}=\nu_{\rm ssa}=5$ GHz and $\nu_{\rm min}=\nu_{\rm max}/30$. 
This model corresponds to a single
zone of a complete inhomogeneous model (e.g., Blandford \& 
K\"{o}nigl \markcite{bland79} 1979).  The low and high energy electrons
are fully mixed.  Assuming equipartition, we find a magnetic field of 0.4
Gauss and a maximum electron Lorentz factor of 60. 
The electron distriubtion chosen here has an equal number of 
low- and high-energy electrons per logarithmic interval.
According to Pacholczyk (1977, Eq. 3.152), such a power law will
contain enough
low-energy electrons near $\gamma_{\min}$ to produce the observed circular
polarization through repolarization.
Intrinsic circular polarization could be as
important as conversion in this model.  The intrinsic synchrotron circular
polarization is given by 
$m_{\rm c}=3\%\,(B/{\rm Gauss})^{1/2}(\nu/{\rm GHz})^{-1/2}=0.9\%$, 
assuming an angle of $60^\circ$
between the magnetic field and the line of sight (Legg \& Westfold 
\markcite{legg68} 1968).  
However,
field reversals and optical depth effects will decrease that number.

The polarization properties of models (e.g., ADAF
and Bondi-Hoyle) that produce gyrosynchrotron
emission with low temperature electrons are largely unexplored
(Narayan \markcite{naray98} \etal 1998, Melia \markcite{melia94} 1994).  
However, Ramaty \markcite{ramat69} (1969) 
did show that circular polarization may
dominate linear polarization in some simple gyrosynchrotron
sources.

Obviously, a more self-consistent treatment of these problems is
required. The circular and linear polarization spectrum will depend on
the electron distribution and the temperature and magnetic field
stratification in the source.  Nevertheless, it seems as if a highly
self-absorbed source with a low high-energy cut-off and a modest
amount of low-energy electrons might explain the observed
properties of Sgr A*.

Observationally, the most crucial steps are the measurement of
the circularly polarized spectrum over a broader frequency range and
its variability characteristics.

This discovery opens a significant new
parameter space for the study of the nearest supermassive black hole
candidate and its environment. This is especially important at centimeter
wavelengths where the morphological structure of Sgr A* will remain
concealed forever due to strong scattering.  In concert with other
radio and millimeter wavelength techniques, we may be closer
to decoding the complex picture of Sgr~A*.

\acknowledgements We thank A.G. Pacholczyk, M. 
Rees, J. Wardle, F. Melia, R. Perley \& R. Beck for many useful discussions.

\newpage

\begin{figure}
\mbox{\psfig{figure=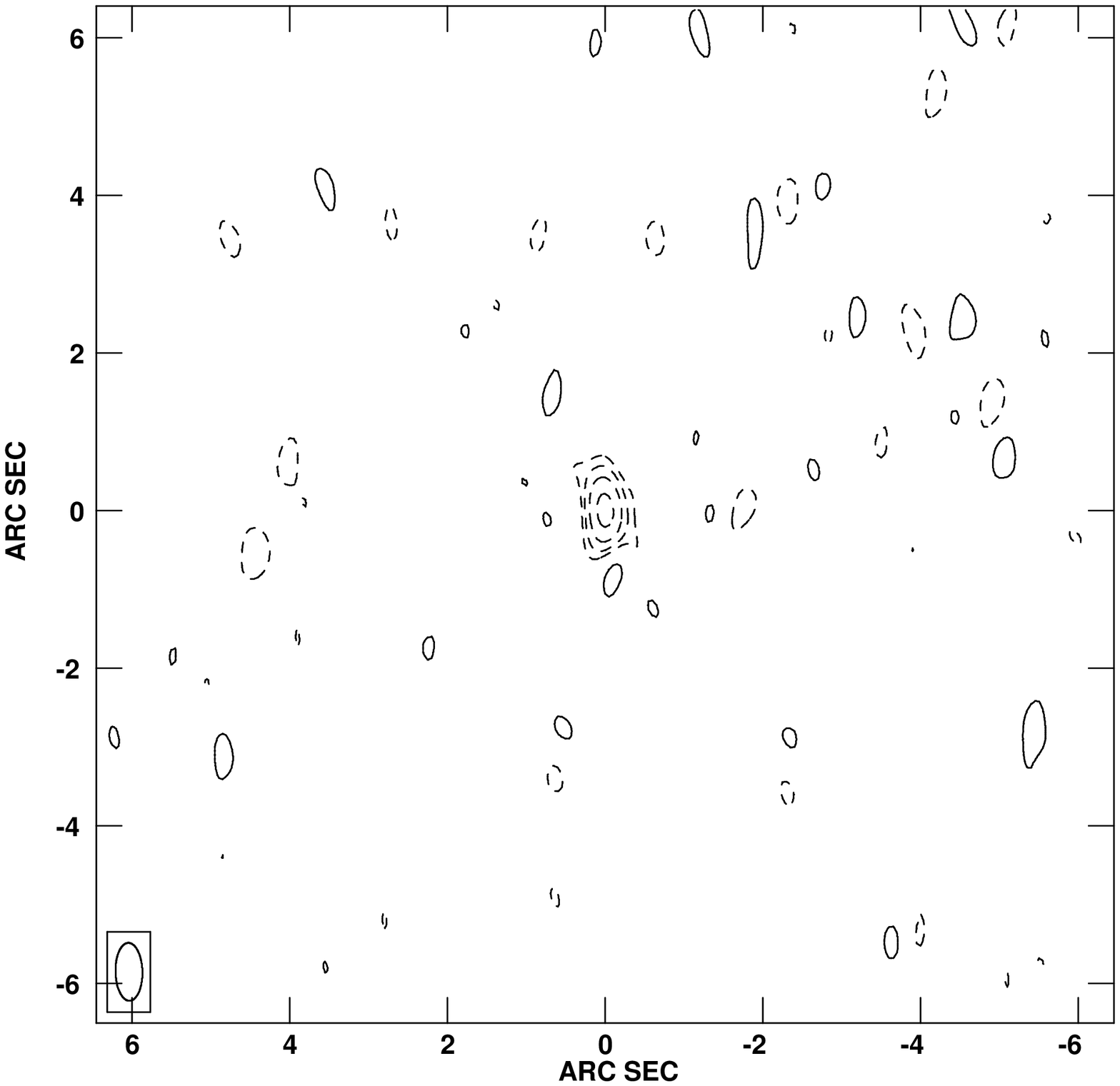,bbllx=36pt,bblly=91pt,bburx=576pt,bbury=701pt,width=\textwidth,angle=270}}
\caption{The Stokes $V$ map for Sgr A* 
on 10 April 1998 at 4.8 GHz.
The peak intensity is -1798 $\mu {\rm Jy\ beam^{-1}}$ .
The rms noise is 68 $\mu {\rm Jy\ beam^{-1}}$.
Contour levels are -16, -8, -4, -2, -1, 1, 2, 4, 8 
and 16 times 175 $\mu {\rm Jy\ beam^{-1}}$.  Negative contours are shown
with dashed lines.}
\end{figure}

\begin{deluxetable}{llrrrrrr}
\tablecaption{Circularly polarized flux at 4.8 GHz and 8.4 GHz}
\tablehead{
                 &                & \multicolumn{3}{c}{4.8 GHz} &
\multicolumn{3}{c}{8.4 GHz} \\
\cline{3-5} \cline{6-8} \\
\colhead{Source} & \colhead{Date} & \colhead{$I$} & \colhead{$P_c$} &
\colhead{$m_c$} &\colhead{$I$} &\colhead{$P_c$} &\colhead{$m_c$} \\
                 &                & \colhead{(Jy)} & \colhead{(mJy)} &
\colhead{(\%)} & \colhead{(Jy)} & \colhead{(mJy)} & \colhead{(\%)} }
\startdata
Sgr~A* & 10 Apr 98 & 0.525 & -1.8  & -0.34 & 0.590 & -1.9 & -0.32 \\
       & 18 Apr 98 & 0.494 & -1.5 & -0.30 & 0.555 & -0.87 & -0.16 \\
       & 24 Apr 98 & 0.611 & -2.6 & -0.42 & 0.738 & -2.1 & -0.29 \\
\cline{2-8}
        & Average & 0.543 & -1.97 & -0.36 & 0.628 & -1.62 & -0.26 \\ 
\\
B1748-253& 10 Apr 98 & 0.488 & 0.26 & 0.05 & 0.278 & -0.23 & -0.09 \\
       & 18 Apr 98 & 0.481 & 0.31 & 0.06 & 0.277 & 0.27 & 0.10 \\
       & 24 Apr 98 & 0.491 & -0.25 & -0.05 & 0.276 & 0.27 & 0.10 \\
\cline{2-8}
        & Average & 0.487 &  0.11 &  0.02 & 0.277 &  0.10 &  0.04 \\ 
\\
B1737-294 & 10 Apr 98 & 0.046 & 0.05 & 0.11 & 0.030 & 0.15 & 0.68 \\
       & 18 Apr 98 & 0.045 & 0.06 & 0.13 & 0.030 & -0.13 & -0.43 \\
       & 24 Apr 98 & 0.046 & 0.07 & 0.15 & 0.029 & 0.10 & -0.34 \\
\cline{2-8}
        & Average & 0.046 &  0.06 &  0.13 & 0.030 &  0.04 &  0.13 \\ 
\\
B1742-283    & 10 Apr 98 & 0.108 & -0.04 & -0.04 & 0.138 & 0.40 & 0.30 \\
       & 18 Apr 98 & 0.105 & 0.08 & 0.07 & 0.136 & 0.27 & 0.20 \\
       & 24 Apr 98 & 0.108 & -0.05 & -0.05 & 0.135 & 0.15 & 0.11 \\
\cline{2-8}
        & Average & 0.107 & -0.00 & -0.00 & 0.136 &  0.27 &  0.20 \\ 
\\
B1745-291   & 10 Apr 98 & 0.101 & -0.13 & -0.13 & 0.096 & 0.14 & 0.15 \\
       & 18 Apr 98 & 0.099 & 0.08 & 0.08 & 0.096 & 0.18 & 0.19 \\
       & 24 Apr 98 & 0.101 & 0.12 & 0.11 & 0.095 & 0.16 & 0.17 \\
\cline{2-8}
        & Average & 0.100 &  0.02 &  0.02 & 0.096 &  0.16 &  0.17 \\ 
\\
B1730-130 & 10 Apr 98 & \dots & \dots &\dots & 5.782 & -1.7 & -0.03 \\
         & 18 Apr 98 & \dots & \dots & \dots & 5.618 & 6.7 & 0.12 \\
         & 24 Apr 98 & \dots & \dots & \dots & 5.467 & 8.1 & 0.15 \\
\cline{2-8}
         & Average & \dots & \dots & \dots & 5.622 & 4.37 & 0.08 \\
\enddata
\end{deluxetable}

\end{document}